\documentclass[12pt]{iopart}

\makeatletter
\def\l@subsubsection#1#2{}
\def\l@subsubsubsection#1#2{}
\makeatother

\usepackage{graphicx,amssymb,amsthm,amsfonts,epsfig,epsf}
\usepackage[usenames]{color}
\usepackage{epstopdf}
\usepackage{aas_macros}
\usepackage{bm}
\usepackage{dcolumn}
\usepackage{latexsym}
\usepackage[linktocpage]{hyperref}
\definecolor{amber(sae/ece)}{rgb}{1.0, 0.49, 0.0}

\def\be{\begin{equation}}
\def\ee{\end{equation}}
\def\beq{\begin{eqnarray}}
\def\eeq{\end{eqnarray}}

\begin{document}

\title{ECO-spotting: looking for extremely compact objects with bosonic fields}

\author{Vitor Cardoso}
\address{CENTRA, Departamento de F\'{\i}sica, Instituto Superior T\'ecnico -- IST, Universidade de Lisboa -- UL,
Avenida Rovisco Pais 1, 1049 Lisboa, Portugal and}
\address{Niels Bohr International Academy, Niels Bohr Institute, Blegdamsvej 17, 2100 Copenhagen, Denmark}

\author{Caio F. B. Macedo}
\address{Faculdade de Física, Universidade Federal do Par\'a, Salin\'opolis, Par\'a, 68721-000 Brazil}

\author{Kei-ichi Maeda}
 \address{Department of Physics, Waseda University, Shinjuku, Tokyo 169-8555, Japan}

\author{Hirotada Okawa}
\address{Waseda Institute for Advanced Study (WIAS), Waseda University, Shinjuku, Tokyo 169-8050, Japan}
		
\begin{abstract}
Black holes are thought to describe the geometry of massive, dark compact objects in the universe. To further support and quantify this long-held belief requires knowledge of possible, if exotic alternatives. Here, we wish to understand how compact can self-gravitating solutions be.
We discuss theories with a well-posed initial value problem, consisting in either a single self-interacting scalar, vector or both. We focus on spherically symmetric solutions,
investigating the influence of self-interacting potentials into the compactness of the solutions, in particular those that allow for flat-spacetime solutions. 
We are able to connect such stars to hairy black hole solutions, which emerge as a zero-mass black hole.
We show that such stars can have light rings, but their compactness is never parametrically close to that of black holes. The challenge of finding black hole mimickers to investigate full numerical-relativity binary setups remains open.
\end{abstract}


\section{Introduction}
Black holes (BHs) occupy a special place among all possible astrophysical objects~\cite{Barack:2018yly,Cardoso:2019rvt,Berti:2015itd}. Once thought as an exotic possibility, they are now acknowledged to be the standard outcome of gravitational collapse. Their growth through accretion and mergers is also tied to the evolution and structure of the host galaxy itself. Most of the BH unique features are related to their compactness, which sets a lower limit on how small heavenly bodies of fixed mass can be. The high compactness of BHs is also responsible for the large gravitational potential at their boundary (the horizon, a ``one-way surface''), for the relativistic motion of matter in their vicinities and therefore for the observational signatures of BHs. Thus, BHs in the universe are identified by looking for massive, dark compact objects.

It is natural to inquire if nature could provide us with astrophysical objects that behave in a similar way to BHs. After all, the standard model of particle physics is able to provide us with an almost endless arrangement of matter. Who knows if standard matter -- together with the dark matter sector -- can conspire to produce other types of massive compact objects~\cite{Cardoso:2019rvt}? The quest for objects with a compactness extremely close to that of BHs is made challenging, and interesting, because of the connection to some fundamental physics issues. In particular, large compactness requires high pressure to sustain large gravitational fields. Large compactness and the absence of horizons also lead to the appearance of stable light rings, which are conjectured to
give rise to nonlinear instabilities~\cite{Cardoso:2019rvt,Keir:2014oka,Cardoso:2014sna,Cunha:2017qtt}. However, in the same way that quantum degeneracy pressure can sustain neutron stars or white dwarfs, it is conceivable that new types of matter also give rise to pressures able to hold them closer to the Schwarzschild radius. In addition, the nonlinear instability is conjectured but not proven, and can be by-passed in different ways, including dissipation mechanisms likely to be present in any astrophysical object~\cite{Cardoso:2014sna}.

The gravitational-wave astronomy era makes the quest for horizonless compact objects more relevant than ever: is there structure close to the Schwarzschild radius, and can we quantify the statement that BHs do exist~\cite{Barack:2018yly,Cardoso:2019rvt,Baibhav:2019rsa}?

The understanding of the above question requires the study of possible compact configurations, their dynamical behavior, formation properties and observational appearance. For isotropic fluid configurations of total mass $M$, the surface of the star sits at an areal radius $R>2.25M$, as imposed by the Buchdahl limit~\cite{PhysRev.116.1027}. It is possible to circumvent this bound, allowing for anisotropies. In fact, in some models one can build configurations with compactness arbitrarily close to that of a BH. Such models include incompressible fluids~\cite{Florides,Raposo:2018rjn}, thin-shell models such as gravastars~\cite{Mazur:2001fv,Mazur:2004fk}, charged configurations~\cite{Lemos:2007yh,Lemos:2020ooh}, among others~\cite{Cardoso:2019rvt}. Most of the previous models are hard to encode in a covariant, fully four-dimensional formulation, within a theory with a well-posed initial value problem. This hinders the investigation of highly dynamical formation scenarios and dynamical properties of such objects.

The above shows the necessity of having controlled simple models that predict highly compact solutions which can be studied with the standard tools available today. Here, we consider self-gravitating configurations built of bosons. We will term these ``boson stars'' (BS), irrespective of whether the boson is a scalar or vector. BSs are stellar-like configurations formed by bounded bosonic fields~\cite{Kaup:1968zz,Liebling:2012fv}, and can come in many flavors, such as complex scalar and vector fields, charged scalar fields configurations, combination of fermionic fields and so on. There are also stellar-like solutions constructed from real bosonic fields, named oscillatons~\cite{Seidel:1991zh,Brito:2015yfh}. The structure of BSs can change drastically from one self-interacting potential to another. There are a plethora of possible self-interacting potentials, with the simplest one being the case in which the potential includes only the mass term.

Among all possible models, the so-called non-topological solitons stand out~\cite{Lee:1986ts,Lee:1991ax}. In this theory, the scalar self-interacting potential can be written as
\begin{equation}
V_S=\mu^2|\Psi|^2\left(1-\frac{|\Psi|^2}{\sigma^2}\right)^2,\label{eq:selfin}
\end{equation}
where $\Psi$ is the complex scalar field and $\sigma$ a constant. The mass parameter $\mu$ is related to the physical boson mass $m_B$ via $m_B=\hbar \,\mu$. 
The above theory is written for a scalar but can be generalized to a vector, as we discuss in Section~\ref{sec:vector}. 
For small enough $\sigma$, false vacuum solutions exist with $|\Psi|\sim \sigma$, and the scalar field can be approximated as a step function, in the so-called thin-wall approximation. 
Solutions with the potential above have been shown to be very compact, enabling the existence of pair of light-rings~\cite{Macedo:2013jja}\footnote{A related, ``axionic'' potential also gives rise to multiple false degenerate  vacua and to double light-rings~\cite{Guerra:2019srj}.}. 
Thus, solutions with false vacua are a good starting point to explore possible limits on the compactness of bosonic field configurations. In this work, we investigate the compactness of BSs with different field content and including higher dimensional setups.

This paper is organized as follows. In Section \ref{sec:neutral} we study complex scalar fields subjected to a potential of the type \ref{eq:selfin} in four and higher dimensions. We focus on the thin-wall approximation, which has been shown to describe well highly compact configurations~\cite{Cardoso:2016oxy,Boskovic:2021nfs}. In Section \ref{sec:vector} we analyze vector boson stars with sextic self-interacting potentials. We describe the influence of the new terms into the compactness of the solutions and explore the possibility of having a thin-wall approximation similar to the scalar field case. In Section~\ref{sec:charged} we explore charged scalar fields, in which the complex scalar is coupled to a real vector field. By focusing on the thin-wall approximation, we see a modest increase of the compactness. We also solve the full nonlinear field equations to show that indeed the thin-wall approximation can be applied in some regimes. Finally, in Section~\ref{sec:conclusion} we present our final thoughts.

\section{Scalar boson stars in $d$ dimensions and the thin wall approximation}\label{sec:neutral}

We start with a theory of a self-interacting, complex scalar field. To be as general as possible, here we consider $d$-dimensional spacetime,
\be
S=\int d^{d}x \sqrt{-g}\left(\frac{R}{16\pi}-g^{\mu\nu}\partial_{\mu}\Psi^*\partial_\nu\Psi-V_S\right)\,,
\ee
where $R$ is the Ricci scalar, and $g$ the metric determinant. We focus on the solitonic potential given in Eq.~(\ref{eq:selfin}), and look for spherically symmetric geometries
\be
ds^2=-e^{v(r)}dt^2+e^{\lambda(r)}dr^2+r^2d\Omega^2_{d-2}\,,
\ee
with $d\Omega^2_{d-2}$ the metric of the unit $(d-2)$ sphere. With the ansatz $\Psi=\phi_0 e^{-i\omega t}$, one finds the equations of motion
%
\beq
&&\left(r^{d-3}e^{-\lambda}\right)'-(d-3)r^{d-4}=-\frac{2r^{d-2}}{d-2}\left(e^{-v}\omega^2\phi_0^2+e^{-\lambda}\phi_0'^2+V_S\right)\,,\\
&&e^{-\lambda}\left[(d-2)rv'+6+d(d-5)\right]-6-d(d-5)\nonumber\\
&&=2r^2\left(e^{-v}\omega^2\phi_0^2+e^{-\lambda}\phi_0'^2-V_S\right)\,,
\label{eq_lambda}
\\
&&\phi_0''+\phi_0'\left(\frac{v'-\lambda'}{2}+\frac{d-2}{r}\right)=\phi_0 e^{\lambda}\left(U_S-\omega^2e^{-v}\right)\,.
\eeq
%
Here, $U_S\equiv \frac{dV_S}{d|\Psi|^2}$ and primes stand for radial derivatives. Minimal (i.e. $\sigma\to\infty$) BSs in $d$-dimensions were analyzed in Ref.~\cite{Blazquez-Salcedo:2019qrz}. Solitonic stars
were studied at length in four-dimensional spacetime. It was shown, both analytically and numerically, that the most compact one have a simple scalar profile, constant in the interior of the object, and with a sharp surface~\cite{Friedberg:1986tq,Kesden:2004qx,Macedo:2013jja}.

We find similar properties in $d-$dimensions. {\it Assuming} that the stars are characterized by a sharp surface and a constant scalar in the interior $\phi_0\approx\sigma$,\footnote{We note that although the scalar field is constant $\phi_0\sim \sigma$ in the interior, this does not mean that we are describing a single physical solution. In fact, when dealing with the full equations, the mass does not increase monotonically with the central scalar field and as such $\phi_0=\sigma$ can actually describe different physical solutions~\cite{Cardoso:2016oxy}.} 
the field equations are simplified considerably~\cite{Friedberg:1986tq}. In this scenario, the function $\lambda$ is discontinuous across the surface\footnote{We note that the full numerical computations show a steep function at the surface, which indicate that the approximation is valid~\cite{Macedo:2013jja}.}. The coupled equations above can be decoupled (since now the scalar is trivial). After manipulating the Einstein's equations, one finds the following equation for the metric function $v$:
\beq
&&(d-3)(d-2) e^{v(r)} \left(v' (d-2+rv') + r v''\right)\nonumber\\
&+& 2 r\sigma^2  \omega^2 \left(-2 (d-3) (d-2) + (5-2d) r v' + r^2 v''\right)=0\,.
\eeq
This equation can be made dimensionless with the variable $x=\sigma\omega r$, after which the quantity $\sigma\omega$ scales out of the problem. Thus, in the interior, our metric function is described by the simple ODE
\beq
&&(d-3)(d-2) e^{v} \left(v' (d-2+xv') + x v''\right)\nonumber\\
&+& 2 x\left(-2 (d-3) (d-2) + (5-2d) x v' + x^2 v''\right)=0\,.\label{ode_solitonic}
\eeq

The integration is performed as follows. We fix a value of $v(x\approx 0)=v_c$ at the center of the star, and then integrate outwards up to the stellar radius $R$, where the interior solution smoothly matches onto a vacuum exterior, described by the Schwarzschild-Tangherlini geometry~\cite{Tangherlini:1963bw}
\be
e^v=e^{-\lambda}=1-\left(\frac{2{\cal M}}{r}\right)^{d-3}\,.\label{tangherlini}
\ee
Here ${\cal M}$ is a mass parameter, related to the spacetime mass $M$ as
\beq
M&=&\frac{(d-2)\Omega_{d-2}}{16\pi G} (2{\cal M})^{d-3}\,,\\
\Omega_{d-2}&=&\frac{2\,\pi^{(d-1)/2}}{\Gamma\left((d-1)/2\right))}\,.
\eeq
The process is a shooting method, such that for a given $v_c$ we shoot for $R$ in which the conditions above are satisfied. Alternatively, one can also prescribe $R$ and shoot for $v_c$. The condition at the origin is equivalent to requiring regularity at the center. Thus, we can compute the mass from Eq.~(\ref{tangherlini}), which is related to the metric function at large $r$ by
\be
\left(2{\cal M}\right)^{d-3}=\frac{v'}{r^{3-d}v'+(d-3)r^{2-d}}\,,
\ee
where a prime stands for derivative with respect to $r$.
Since $v'$ is continuous at the surface, the mass is evaluated by the surface values 
of the inside metric components.

\begin{figure}
	\centering	
	\includegraphics[width=0.55\linewidth]{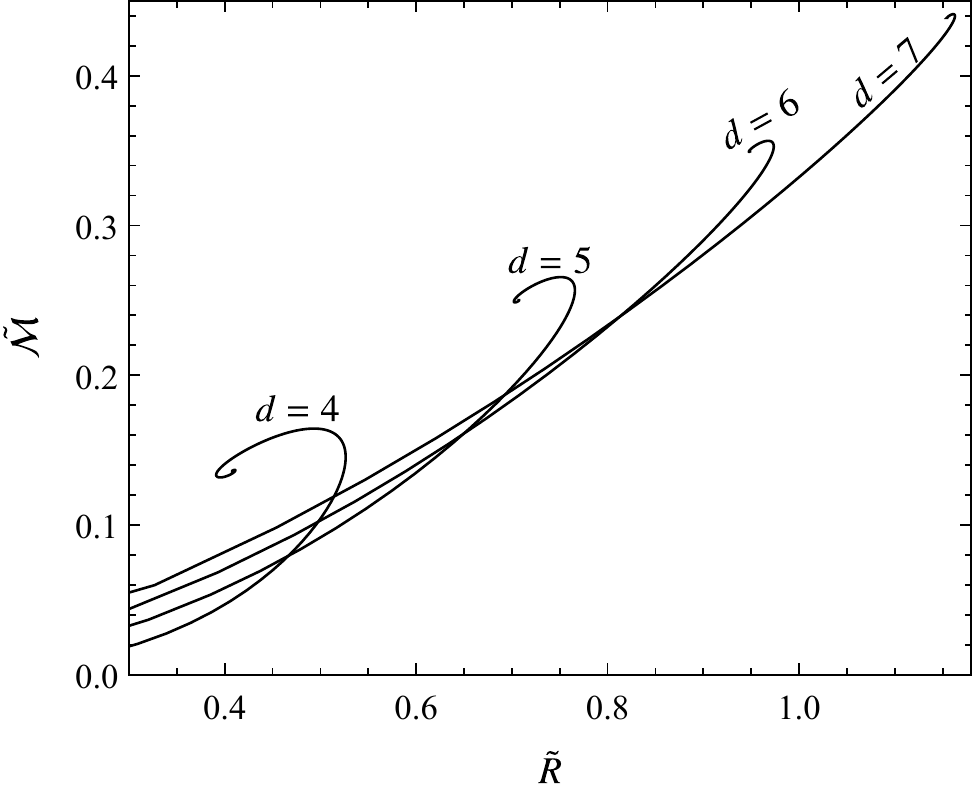}
	\caption{Mass-radius curve for d-dimensional scalar BSs in the thin-wall approximation. Units are such that $\sigma \omega=1$. For higher dimensions, the curve is approximately a straight line for intermediate to high values of the mass.}
	\label{fig:mr_d}
\end{figure}

We followed the above procedure to integrate the field equation~(\ref{ode_solitonic}) for different values of $v_c$ and different spacetime dimension $d$, constructing the mass-radius curves for the BS solutions. These are important for a stability analysis but also because they inform us on the compactness of such configurations.
The results are shown in Fig.~\ref{fig:mr_d}, where we plot the mass-radius curve using dimensionless quantities ${\tilde{R}}=\sigma \omega R$ and  ${\tilde{\cal M}}=\sigma \omega {\cal M}$. The qualitative behavior at any $d$ is essentially the same, with the mass increasing with the radius, reaching the marginally stable configuration (represented by the maximum mass configuration\footnote{Although we do not prove that the maximum mass configuration is marginally stable, for many different BS configurations this is indeed true, so we adopt this as a conjecture at this stage.}), then inspiralling back in a sequence of unstable solutions. In the higher dimensional case, the mass-radius curve approaches a straight line, indicating that the compactness is approximately a constant when fixing the dimension $d(\gg4)$ of the spacetime. Notice that already at a moderately large value of $d=7$ the trend is apparent.

\begin{figure}
\centering	\includegraphics[width=0.55\linewidth]{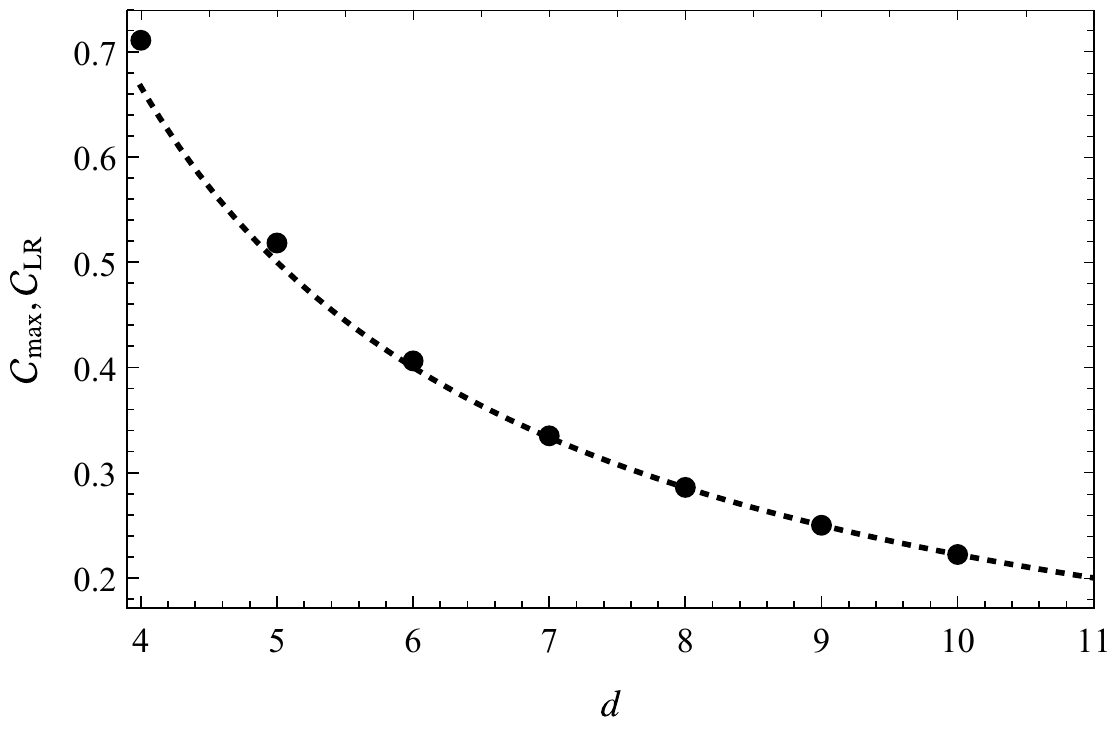}
\caption{Comparison between the maximum compactness configurations of the $d$-dimensional BS in the thin wall approximation (black dots) with the compactness of the light-ring (dotted line).}\label{fig:compactd}
\end{figure}
We define the compactness of the object as 
\be
{\cal C}=\left(\frac{2{\cal M}}{R}\right)^{d-3}=\frac{xv'}{d-3+xv'}\Big{|}_{x=x_S}\,,
\ee
with primes now standing for derivative with respect to $x$, and $x_S$ is the adimensional surface radius.

We are now able to formulate the question we wish to address here: {\it what is the maximum compactness of the object once the radius and $v(x=0)$ are allowed to vary?} Notice that the maximum compactness solution will not necessarily be in the stable branch. By analyzing the solutions explored in Fig.~\ref{fig:mr_d}, we find that the maximum compactness for $d=4$ is given by
\be
{\cal C}_{\rm max}=0.7108\,,\quad d=4\,.
\ee
This value describes surprisingly well the numerical results obtained from solving the full equations without relying on the thin-wall approximation~\cite{Macedo:2013jja,Cardoso:2016oxy}. Regarding the  maximum compactness in higher dimensions, we notice that when $d\to \infty$ the maximum compactness is well described, numerically, by that of a surface at the light ring. For a Tangherlini BH, the light ring lies at $r_{\rm LR}^{d-3}=(d-1){\cal M}^{d-3}$~\cite{Cardoso:2008bp}, corresponding to a critical compactness
\be
{\cal C}_{\rm LR}=\frac{2}{d-1}\,.
\ee
This value describes our numerical results to better than $0.1\%$ when $d>7$, improving the agreement even more for higher dimensions. In Fig.~\ref{fig:compactd} we show the value of the maximum compactness of the higher dimensional BS solutions with the light-ring.

The above result can be found analytically by solving the differential equations in the large $d$ limit. In fact, one can show that such steep profiled stars must end at the light ring. For large $d$, Eq.~(\ref{ode_solitonic}) is simplified to
\be
d e^{v}v'-4x=0\,,
\ee
which is solved by 
\be
v=\log(k+2x^2/d)\,,
\ee
with $k>0$ to ensure a real geometry at the origin. We then find the compactness
\be
{\cal C}=\frac{2}{d-1+d(d-3)k/(2x_S^2)}\leq \frac{2}{d-1}\,.
\ee
This result describes very well the numerical solution of ODE (\ref{ode_solitonic}).

As we saw there is good evidence that scalar-field theories where the field has two minima to decay into will never have radius smaller than $R\sim2.81M$ (for $d=4$). Does this mean that there is no hope of achieving higher compactness configurations when a single scalar field is present? As we also mentioned, axionic potentials can allow for multiple false vacua solutions, having multiple minima~\cite{Guerra:2019srj}. While the scalar field in the axionic case can still be represented by a constant in a region near the center of the star, depending on the value on the parameters, the scalar field may show a multiple step-function behavior, with multiple thin walls. This could be an interesting venue in looking for compact solutions with smooth potentials.

Having established the limits for scalar BSs described by the thin wall approximation, we now explore other field content.

\section{Vector boson stars with self-interactions\label{sec:vector}}
\subsection{Equations and boundary conditions}
Consider a theory describing a complex vector field with self-interactions, in four-dimensional spacetime. We here extend some results discussed in Refs.~\cite{Brito:2015pxa,Loginov:2015rya,Minamitsuji:2018kof}. The action is described by 
\begin{equation}
S=\int d^4x \sqrt{-g}\left[\frac{R}{16\pi}-\frac{1}{4}F_{\mu\nu}\bar{F}^{\mu\nu}-\frac{1}{2}V(|A|^2)\right]\,,
\end{equation}
where $|A|^2=A_\mu\bar{A}^\mu$, and  $F_{\mu\nu} \equiv \nabla_{\mu}A_{\nu} - \nabla_{\nu} A_{\mu}$. We take the following self-interacting potential,
\begin{equation}
V(|A|^2)=\mu^2|A|^2+\frac{a}{2}|A|^4+\frac{b}{3}|A|^6\,.	\label{eq:potential_vec}
\end{equation}
Note that a potential similar to the solitonic potential in Eq.~(\ref{eq:selfin}) can be recovered for particular choices of the constants $(a,b)$. We explore this possibility later on. The equations of motion of the above theory are given by
\begin{eqnarray}
	G_{\mu\nu}&= 8\pi T_{\mu\nu},\label{eq:einstein}\\
	\nabla_\mu F^{\mu\nu}&=UA^\nu,\label{eq:vector}
\end{eqnarray}
where $U=dV(|A|^2)/d|A|^2$, and the stress-energy tensor is
\begin{eqnarray}
T_{\mu\nu}&=\frac{1}{2}\left(F_{\mu\alpha}{\bar{F}_\nu}^\alpha+\bar{F}_{\mu\alpha}{F_\nu}^\alpha\right)-\frac{1}{4}g_{\mu\nu}|F|^2+U A_{(\mu}\bar{A}_{\nu)}-\frac{1}{2}g_{\mu\nu}V(|A|^2)\,.
\end{eqnarray}
We note that the Lorentz condition is no longer compatible with the equations of motion. Instead, due to the self-interaction terms, from Eq.~(\ref{eq:vector}) the following constraint
needs to be satisfied
\begin{equation}
\nabla_{\mu}\left[UA^{\mu}\right]=0\,.	\label{eq:constraint}
\end{equation}
This constraint reduces to the Lorentz condition when $V=\mu^2 |A|^2$. The theory also has a conserved current associated with the $U(1)$ symmetry $A^\mu\to e^{i\alpha}A^\mu$, given by
\begin{equation}
j^{\mu}=\frac{i}{2}\left(\bar{F}^{\mu\nu}A_\nu-F^{\mu\nu}\bar{A}_\nu\right).
\end{equation}

We want to analyze self-gravitating solitons in the above theory. Following Ref.~\cite{Brito:2015pxa}, it is useful to work with the following ansatz for the metric and vector potential
\begin{eqnarray}
ds^2&=-F_m^2B(r)dt^2+B^{-1}dr^2+r^2d\Omega^2\,,\\
\mathbf{A}&=e^{-i\omega t}\left[fdt+i gdr\right]\,,
\end{eqnarray}
where $f,\,F_m$ and $g\,, B$ are functions of the radial coordinate $r$ only.
From the Einstein field equations~(\ref{eq:einstein}), we obtain a system of two first-order equations for the metric functions
\begin{eqnarray}
m'&=&\frac{2 \pi  r \left({B} r \left({F_m}^2 {V}+({f'}-{g} \omega )^2\right)+2 {f}^2 r {U}\right)}{{B} {F_m}^2},\label{eq:derm}\\
F_m'&=&\frac{4 \pi  r {U}}{{F_m}} \left(\frac{{f}^2}{{B}^2}+{F_m}^2 {g}^2\right).\label{eq:derf}
\end{eqnarray}
The equations for the vector field are 
\begin{equation}
	{g}-{f'} \omega=\frac{{B} {F_m}^2 {g} {U}}{\omega}\,,\qquad~\frac{d}{dr}\left[\frac{r^2(f'-\omega g)}{F_m}\right]=\frac{U r^2 f}{F_m B}\,.\label{eq:derpot}
\end{equation}
These equations reduce to the ones in Ref.~\cite{Brito:2015pxa} when $V=\mu |A|^2$. The above equations imply the constraint given by Eq.~(\ref{eq:constraint}).

We need to supplement the differential equations with boundary conditions. Instead of using $f$ and $g$ to describe the vector potential, we use instead
\be
\tilde f = \frac{f}{F_mB^{1/2}}\,,\qquad \tilde g=Bg\,.
\ee
In this way, $|A|^2=-\tilde{f}^2+\tilde{g}^2$. We require regularity at the center of the BS and asymptotic flatness.
From the condition of regularity at the origin, we find the behavior
\beq
F_m^2B&\approx& a_0+{\cal O}(r^2)\,,\quad B\approx 1+{\cal O}(r^2)\,,\\
\tilde{f}&\approx& \tilde{f}_c+{\cal O}(r^2)\,,\quad \tilde{g}\approx -\frac{\omega \tilde{f}_c}{3\sqrt{a_0}}r+{\cal O}(r^2)\,,
\eeq
where ${f}_c$ and $a_0$ are constants. From the above we see that $|A|^2$ is negative near the center of the star. The equations of motion (\ref{eq:derm})--(\ref{eq:derpot}) are invariant under the transformation
\begin{equation}
F_m\to \alpha F_m\,,\qquad f\to \alpha f\,,\qquad \omega\to \alpha \omega\,.
\end{equation}
We can use this property to set $a_0=1$ in the integrations. This will lead to a solution which is not in the standard Schwarzschild coordinates, but we can always rescale back after the integration, such that $F_m^2B \to 1$ at  infinity.

Not all values of $\tilde{f}_c$ allow for solitonic configurations~\cite{Minamitsuji:2018kof}. By a direct manipulation of the equations of motion, we obtain that the first derivative of $g$ diverges if
\begin{equation}
{\cal H}=\mu^2 + b \tilde{f}^4 + 3 a \tilde{g}^2 + 5 b \tilde{g}^4 - \tilde{f}^2 (a + 6 b \tilde{g}^2)\,,
\end{equation}
vanishes. By expanding the equations near the origin, this translates to
\begin{equation}
{\cal H}=\mu^2-a \tilde{f}_c^2+b\tilde{f}_c^4\,.
\end{equation}
from which we can find the limits for $\tilde f_c$. The roots of the above equations are
\begin{equation}
\tilde{f}_c=\sqrt{\frac{a\pm\sqrt{a^2-4b\mu^2}}{2b}}, ~{\rm and}~
\tilde{f}_c=-\sqrt{\frac{a\pm\sqrt{a^2-4b\mu^2}}{2b}}.\label{eq:cond}
\end{equation}
If these roots are not real, solitons could, in principle, exist for all real values of $\tilde f _c$. 
From Eq.~(\ref{eq:cond}), the condition to avoid the singular behavior at the center in $g'(r)$, therefore, reads $a^2<4b\mu^2$. The exact same condition holds for solitons in flat spacetime with the same potential~\cite{Loginov:2015rya}. As such, flat-spacetime counterparts to solitonic configurations exist only for $a^2<4b\mu^2$ (i.e., for real values of $\tilde f _c$). Note, however, that the above condition is only an \textit{indication} for the existence of gravitating solitons. Even with no real solution $\tilde{f}_c$ to Eq.~(\ref{eq:cond}), there might be singularities in a point $r>0$ within the star, due to higher order corrections in ${\cal H}$, for some $\tilde{f}_c$~\cite{Minamitsuji:2018kof}. In fact, in our computations with $a^2<4b\mu^2$ we always found a configuration in which $g'(r)$ diverges for a finite $r$, and we show one of these configurations below. We also note that the condition (\ref{eq:cond}) is dependent of the self-interaction and, as such, other types of potential could heal such behavior.

At large spatial distances, we require asymptotic flatness, which implies
\be
F_m\to 1\,,\quad	    B\to 1-2M/r\,,\quad     \tilde{f}\to0\,,\,\tilde{g}\to 0\,,\quad     {\rm when}\,\, r\to \infty\,.
\ee
We can also extract the spacetime mass $M$ from this asymptotic behavior, as in the previous section. There are many possible definitions of stellar radius for BSs. Here, we define the radius $R$ such that  $99\%$ of the mass is inside $R$, i.e., $m(R)=0.99M$. 
 We use units such that $\mu=1$ in the integrations. This effectively sets a length scale for the variables.

\subsection{Solutions and their compactness}
\begin{figure}
	\includegraphics[width=.49\linewidth]{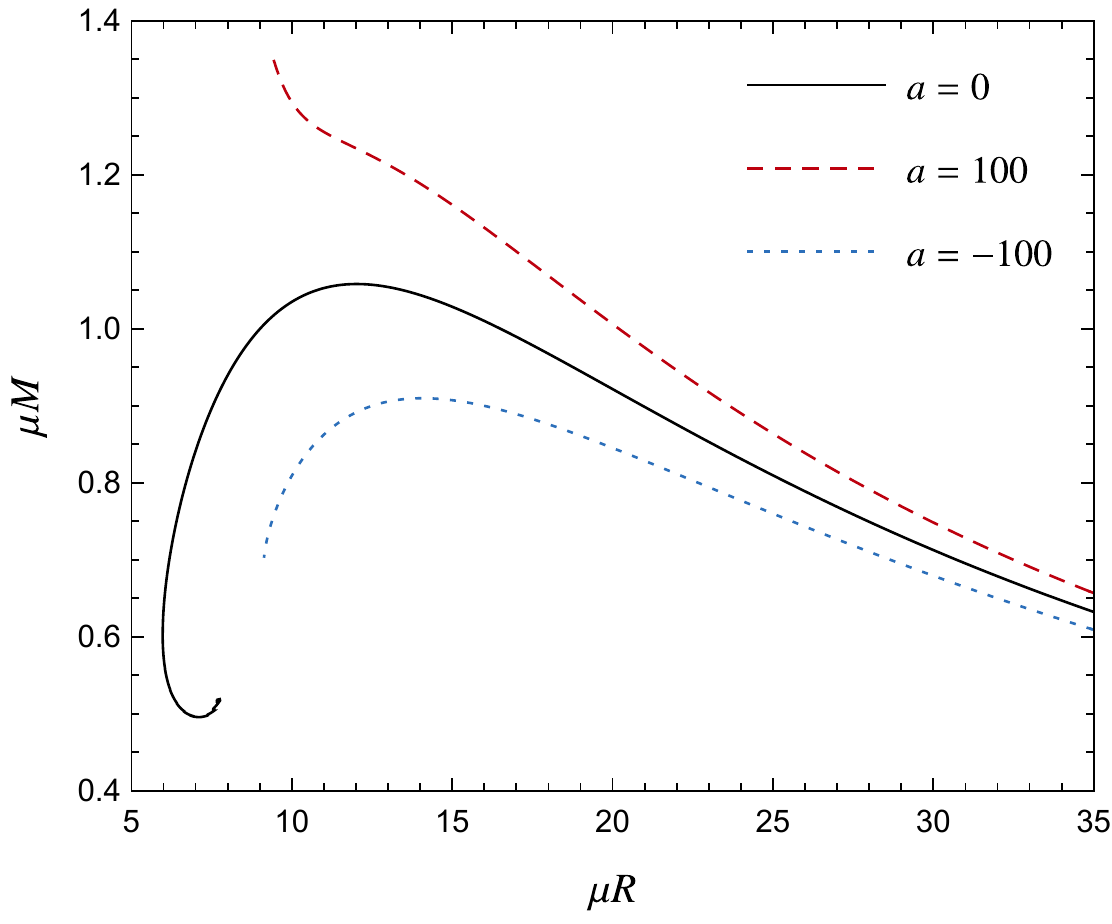}
	\includegraphics[width=.49\linewidth]{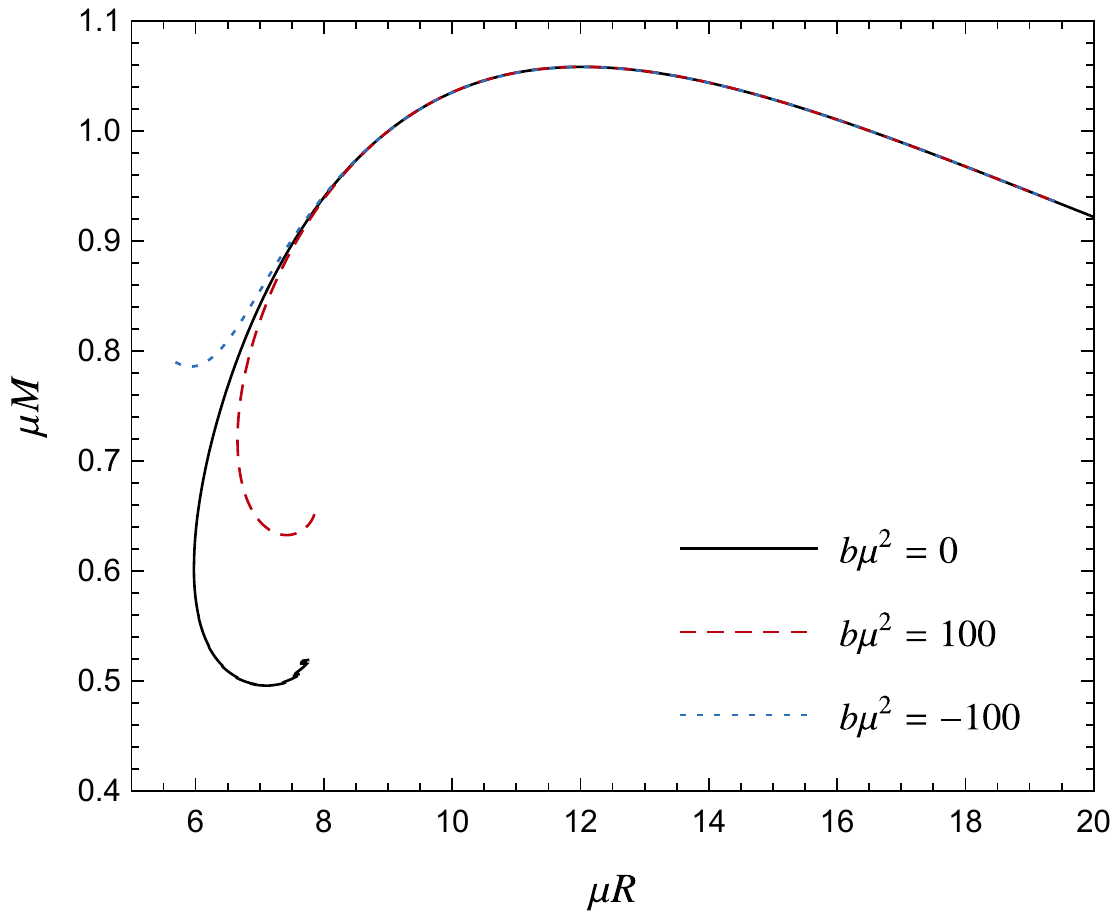}
	\caption{Mass-radius relation for vector BSs with $b=0$ (left panel) and $a=0$ (right panel). Positive values of $a$ lead, in general, to solutions with higher compactness. The curves with self-interactions ($a\neq 0$) in the potential come to a stop at the point in which a singularity is found in the integration, related to the divergence of $g'(r)$. Notice that solutions with $b=0$ were termed Proca stars and described previously~\cite{Brito:2015pxa}.}
	\label{fig:individual}
\end{figure}
\begin{figure}
	\centering\includegraphics[width=.5\linewidth]{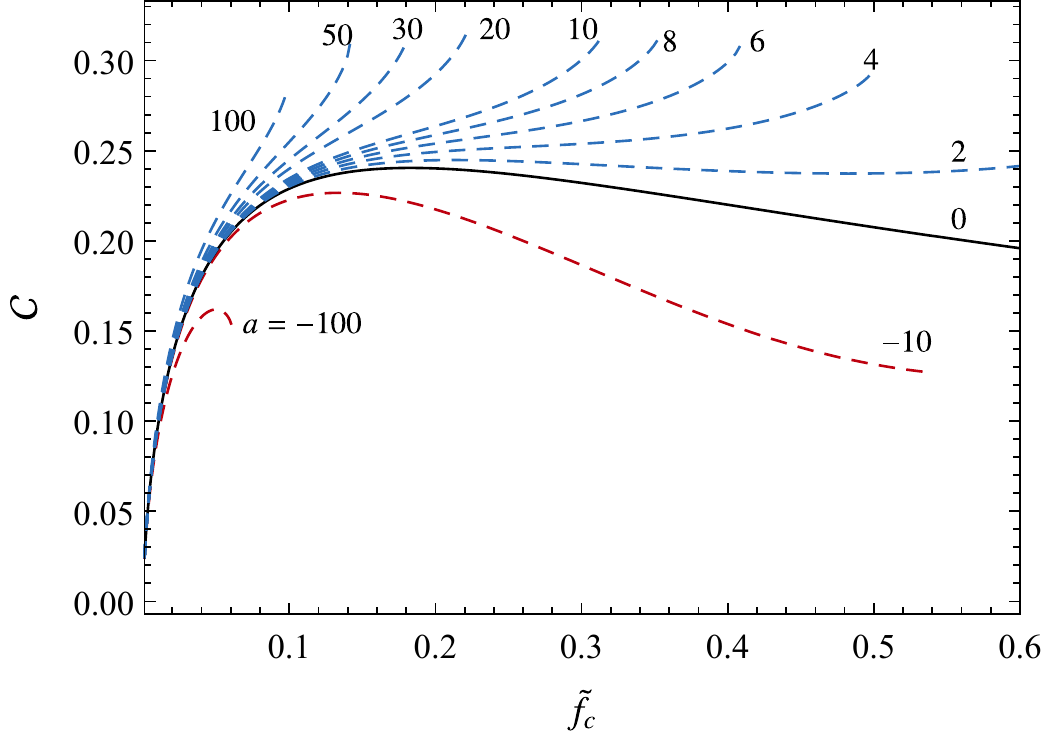}
	\caption{Compactness of vector BSs, neglecting sixth order terms in the potential. The curves and at the place the solutions become singular, which  for the case of positive $a$ is ${\tilde f}_c=\mu/\sqrt{a}$. For negative $a$, as noted in Ref.~\cite{Minamitsuji:2018kof}, although the solutions do not diverge at the origin, they eventually do for a finite value of $r$. For the cases with quartic potential, we have ${\cal C}_{\rm max}\approx0.316$.}\label{fig:asummary}
\end{figure}
Figure~\ref{fig:individual} shows the mass-radius diagram for these BSs, for selected values of the constants $(a,\, b)$. The curve for $a=b=0$ agrees with the results of Ref.~\cite{Brito:2015pxa}. Solutions with only quartic interactions ($b=0$) have higher compactness for larger (positive) values of $a$, and agree with the results of Ref.~\cite{Minamitsuji:2018kof}. The curves stop at the critical value of $\tilde{f}_c$, for which the solution has a singularity in the differential equations at some value of $r$, as explained previously. We also note (cf. right panel of Fig.~\ref{fig:individual}) that positive values of $b$ bend the mass-radius curve to the left. This behavior, together with the fact that flat spacetime solutions exist for $a^2<4b \mu^2$, suggests that there could exist combinations $(a,b)$ which could provide the necessary ingredient to heal the possible divergences in $g'(r)$ to complete the mass-radius curves. However, as we shall see below, other problems related to the divergence of $g'(r)$ appear.

The compactness of some of these BS solutions with $b=0$ are shown in Fig.~\ref{fig:asummary}. We see that the compactness increases with $a$, and ${\cal C}_{\rm max}$ already has its peak around $a\sim 10$, with ${\cal C}_{\rm max}\approx 0.316$. As such, we can fix a value for $a$, investigating the effect of $b$, varying about the point in which flat spacetime solutions start to exist, i.e $b=b_c=a^2/(4\mu^2)$, but the conclusion is similar for other cases as far as we could verify.

\begin{figure}
	\centering\includegraphics[width=.5\linewidth]{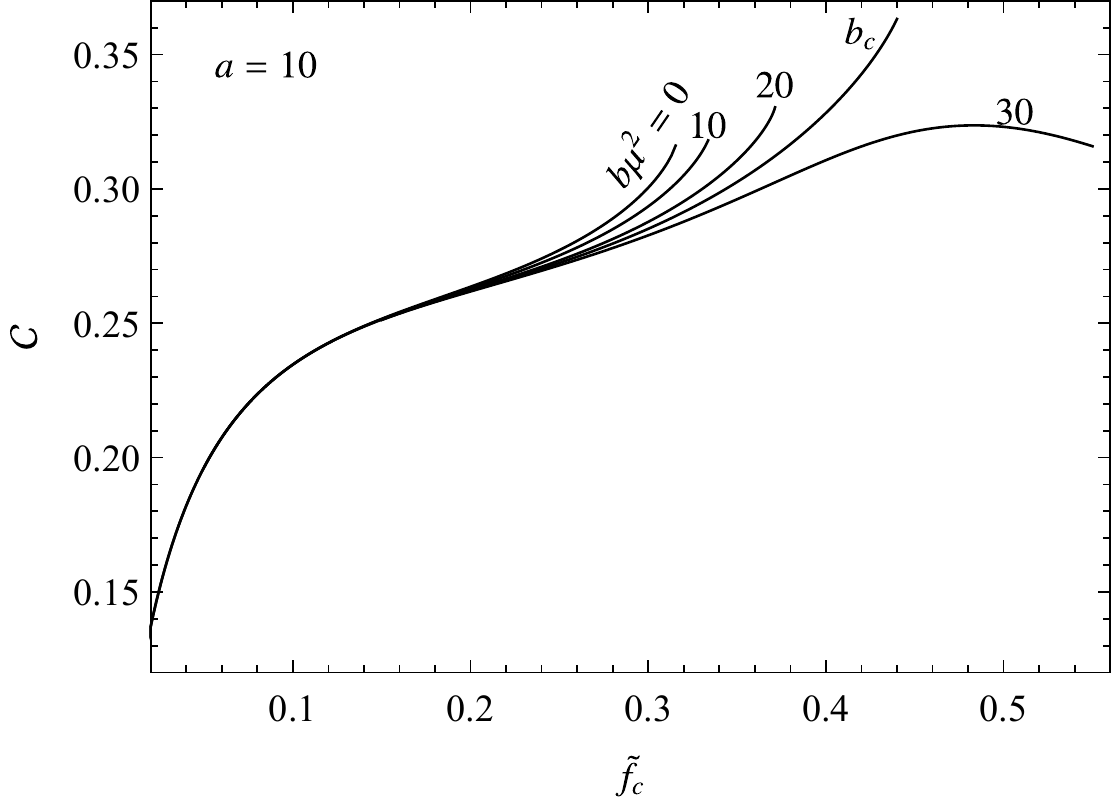}\includegraphics[width=.5\linewidth]{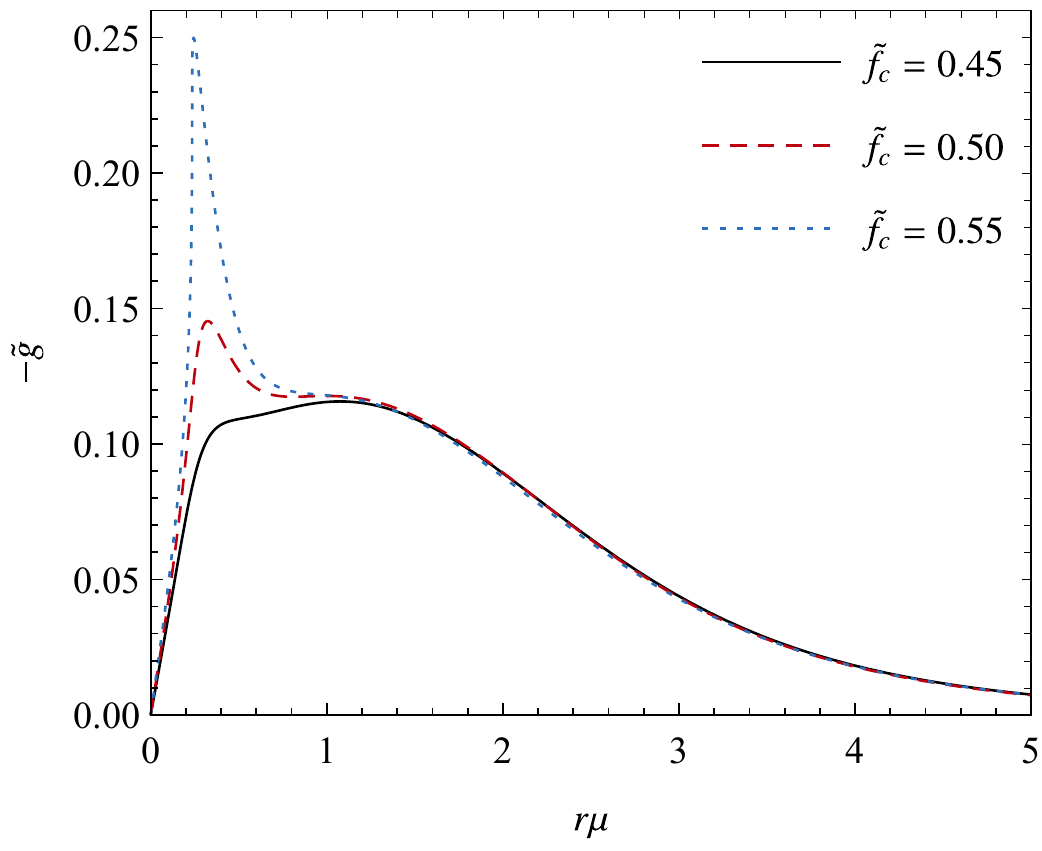}
		\caption{Compactness of self-interacting vector BSs, considering $a=10$ and different values for parameters $b$ (left panel). We note that even though the condition (\ref{eq:cond}) indicates that there is no divergence for $b>b_c$ ($b_c\mu^2=25$ in this case) at the center of the star, $g'(r)$ still diverges at a finite $r$, beyond which we do not find any solutions, which is illustrated in the right panel for the case $(a,b\mu^2)=(10,30)$, for which the critical $\tilde{f}_c$ is not real. The maximum compactness in this case is ${\cal C}_{\rm max}\approx0.363$.}\label{fig:veccom}
\end{figure}

In Fig.~\ref{fig:veccom} we consider $a=10$. As we increase $b$, the point at which the solutions cease to exist move to larger ${\tilde f}_c$, as predicted by condition~(\ref{eq:cond}). For $b>b_c$, although we do not have singularities at the center, $g'(r)$ becomes singular at some finite radius within the star, at which the solutions terminate. In the right panel of Fig.~\ref{fig:veccom} we show $\tilde g$ for some BSs close to this critical point. We can see that the value of $b$ has little impact into the compactness of the solutions but can, however, stabilize some of them. The maximum compactness in this case is ${\cal C}_{\rm max}\approx 0.363$.

\subsection{The possibility of degenerate vacuum and the thin-wall approximation}
The potential described by Eq.~(\ref{eq:potential_vec}) has the possibility of generating false vacua for a given value of the vector $A^\mu\neq0$, i.e. $V(|A|^2)=U(|A|^2)=0$. Inspired by the scalar case, one may be tempted to search near false vacua solitons, given that in the scalar case they provide highly compact configurations. We search for values of $(a,b)$ such that the potential can be written in the following form
\begin{equation}
V(|A|^2)=\mu^2|A|^2\left(1+\frac{|A|^2}{\sigma^2}\right)^2.
\label{eq:vac_potential}
\end{equation}
Given that near the origin $|A|^2\sim-\tilde{f}_c^2$, we have that the potential vanishes for $\tilde{f}_c=\sigma$. By comparing the above potential with Eq.~(\ref{eq:potential_vec}), we obtain that we can map one into the other by choosing the following values for $(a,b)$,
\begin{equation}
a=4\mu^2 /\sigma^2,b=3\mu^2 /\sigma^4.
\end{equation}
Note, however, that for these values of $(a,b)$ we have the following critical points related to divergences at the origin [cf. Eq.~(\ref{eq:cond})]
\begin{equation}
|\tilde{f}_c|=\sigma\{1/\sqrt{3},1\}.
\end{equation}
Therefore, in constructing successive solutions of vector BS to build the mass-radius curve, we eventually reach $\tilde{f}_c=\sigma/\sqrt{3}$, which generate a divergent behavior in $g'(r)$. This indicates either an interruption of the solutions, as some of the cases presented previously,  or a discontinuity in the mass-radius relation. We searched for solutions considering $\tilde{f}_c>\sigma/\sqrt{3}$ for some particular cases, finding none. We argue that false vacuum solutions in a similar fashion of the scalar case does not exist in the vector case.

We note, however, that thin-wall regimes might exist in the vector BS case for the potential~(\ref{eq:potential_vec}). It was shown that in the flat spacetime case~\cite{Loginov:2015rya} this indeed is the case. However, the thin-wall regime in flat spacetime is more closely related to a near constant value of $g(r)$ within the star, instead of $f(r)$.  In addition, it is not clear whether the divergence of $g'(r)$ for finite $r$ also happens in the flat spacetime counterpart, as this type of divergence usually occurs for high values of $\tilde{f}_c$, a regime in which gravity matters. Further studies are needed to explore self-interactions in the vector case to compare the cases with and without the presence of gravity, as well as study ways to construct solutions for all ${\tilde f}_c$.

\section{Scalars and vectors: charged boson stars}\label{sec:charged}
Finally, we consider an action involving one complex, charged massive scalar $\Psi$ with self-interacting potential $V_S$ and a massless vector
field $A_{\mu}$,
\beq
S&=&\int d^4x \sqrt{-g} \bigg( \frac{R}{\kappa}-\frac{1}{4}F^{\mu\nu}F_{\mu\nu}-\frac{1}{2}g^{\mu\nu}\Psi^{\ast}_{,\mu}\Psi^{}_{,\nu}-\frac{1}{2}V_S(|\Psi|^2)\nonumber\\
&+&i\frac{q}{2}A_{\mu}\left(\Psi\nabla^{\mu}\Psi^{\ast}-\Psi^{\ast}\nabla^{\mu}\Psi\right)-\frac{q^2}{2}A_{\mu}A^{\mu}\Psi\Psi^{\ast}\bigg) \, ,\label{eq:MFaction}
\eeq
where $\kappa=16\pi$.
The resulting equations of motion are
%
\begin{eqnarray}
&&\left(\nabla_{\mu}-iqA_{\mu}\right)\left(\nabla^{\mu}-iqA^{\mu}\right)\Psi =\frac{d V_s}{d |\Psi|^2}\Psi	\,,	\label{eq:MFEoMScalar}\\
&&\nabla_{\mu} F^{\mu\nu} = q^2\Psi\Psi^{\ast}A^{\nu}-i\frac{q}{2}\left(\Psi\nabla^{\nu}\Psi^{\ast}-\Psi^{\ast}\nabla^{\nu}\Psi\right) \,,\label{eq:MFEoMVector}\\
&&\frac{1}{\kappa} \left(R^{\mu \nu} - \frac{1}{2}g^{\mu\nu}R+\Lambda g^{\mu\nu}\right) =
- \frac{1}{8}F^{\alpha\beta}F_{\alpha\beta}g^{\mu\nu}+\frac{1}{2}F^{\mu}_{\,\,\alpha}F^{\nu\alpha}\nonumber\\
&-&\frac{1}{4}g^{\mu\nu}\left( \Psi^{\ast}_{,\alpha}\Psi^{,\alpha}+V_S\right)
+\frac{1}{4}\left( \Psi^{\ast,\mu}\Psi^{,\nu}+ \Psi^{,\mu}\Psi^{\ast,\nu} \right)\nonumber\\
&-&i\frac{q}{2}A^{\mu}\left(\Psi\nabla^{\nu}\Psi^{\ast}-\Psi^{\ast}\nabla^{\nu}\Psi\right) -\frac{q^2}{4}g^{\mu\nu}\Psi\Psi^{\ast}A_{\alpha}A^{\alpha}\nonumber\\
&+&\frac{q^2}{2} \Psi\Psi^{\ast}A^{\mu}A^{\nu}+i\frac{q}{4}g^{\mu\nu}A_{\alpha}\left(\Psi\nabla^{\alpha}\Psi^{\ast}-\Psi^{\ast}\nabla^{\alpha}\Psi\right)\,.\label{eq:MFEoMTensor}
\end{eqnarray}
%
We consider spherically symmetric spacetimes. We use the ansatz
\begin{eqnarray}
ds^2&=&-e^{v(r)}dt^2+e^{\lambda(r)}dr^2+r^2d\Omega^2,\\
\Psi&=&\phi_0(r)e^{-i \omega t},\\
A_\mu&=&\{A_0(r),0,0,0\}.
\end{eqnarray}
By substituting the above ansatz into Eqs.~(\ref{eq:MFEoMScalar})--(\ref{eq:MFEoMTensor}), we obtain the following differential equations
%
\beq
\lambda '+\frac{e^{\lambda }-1}{r}=\frac{1}{2}  \kappa  r e^{-v} \left(A_0'^2+2 e^{\lambda } \phi_0^2 (q A_0+\omega)^2+2 e^{v} \left(e^{\lambda } V_S+\phi_0'^2\right)\right)\label{eq:einl}\,,\\
v'+\frac{1-e^{\lambda }}{r}=-\frac{1}{2}  \kappa  r e^{-v} \left(A_0'^2+2 e^{\lambda } \left(e^{v} V_S-\phi_0^2 (q A_0+\omega)^2\right)-2 e^{v} \phi_0'^2\right)\,,\\
A_0''+\frac{\left(4-r \left(\lambda '+v'\right)\right)}{2 r}A_0' -2 q e^{\lambda } \phi_0^2 (q A_0+\omega)=0\,,\\
\phi_0''+ \left(\frac{v'-\lambda '}{2}+\frac{2}{r}\right)\phi_0' +\phi_0 e^{\lambda -v} \left((q A_0+\omega)^2-e^{v} U_S\right)=0\label{eq:einphi}\,.
\eeq
In the following, we solve the above system in two different regimes.

\subsection{Thin-wall approximation}
%
\begin{figure}
	\centering\includegraphics[width=.5\linewidth]{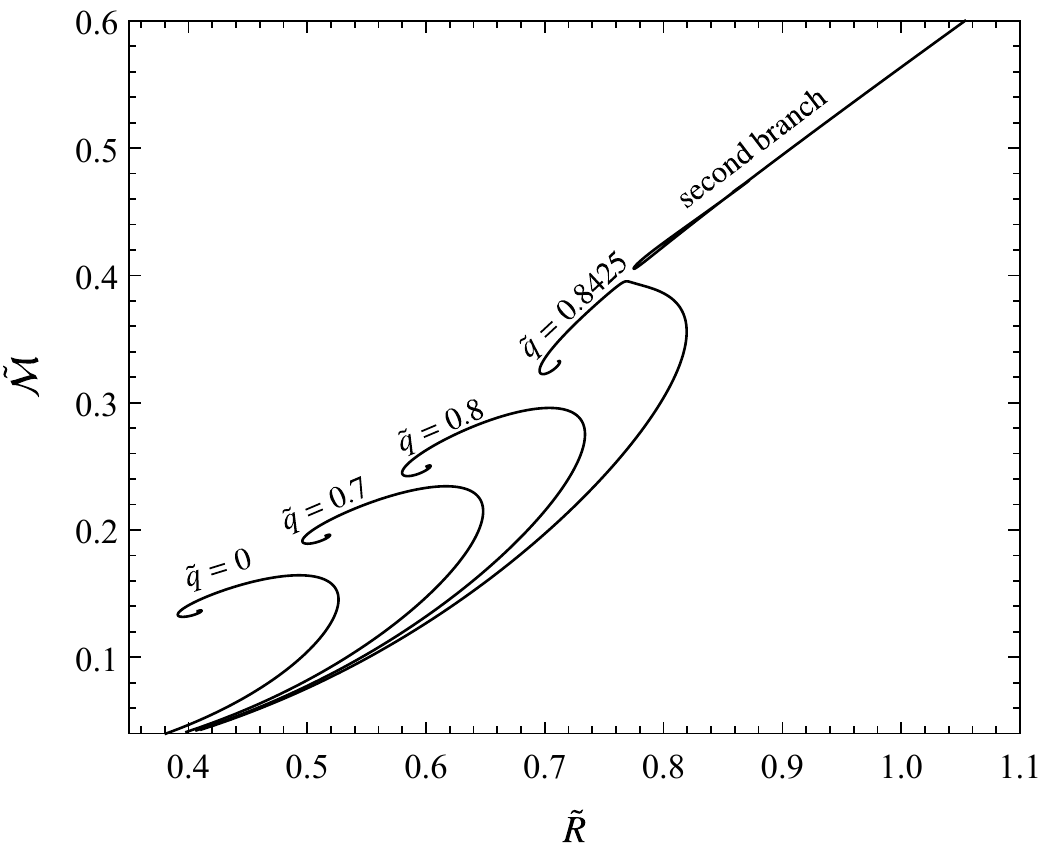}
	\caption{Mass-radius for charged solitonic stars. Tilded quantities refers to the dimensionless variables, which scale in the same way as in Sec.~\ref{sec:neutral}. Note that, as in the standard charged case, the maximum mass increases with the field charge $q$. The second branch refers to solutions that appear in the solitonic case when the field is charged, for which $({\tilde R},{\tilde M})$ quickly increase as $q$ decreases. }
	\label{fig:mass-radius-charged}
\end{figure}
By assuming that the scalar field has a constant value in the interior of the star $\phi_0(r)\sim \sigma$, followed by a sharp decrease at the surface, and that this constant value describes a false vacuum, we can search for solutions as in Sec.~\ref{sec:neutral}. We can greatly simplify the equations. We can write the equations in units of $16\pi G=1$ and rescaling the metric $e^{v}$, the potential $A_0$ and the charge as
\be
e^{v}\to \sigma^2\omega^2e^{v}\,,\qquad A_0\to \sigma \omega A_0\,,\qquad  q \to  \frac{q}{\sigma}\,.
\ee
they take the following form
\begin{eqnarray}
&&\frac{1}{2} x  e^{-v} A_0'^2-x (q A_0+1)^2 e^{\lambda -v}+\frac{1-e^{\lambda }}{x}+v'=0\,,\\
&&-\frac{1}{2} x e^{-v} A_0'^2-x (q A_0+1)^2 e^{\lambda -v}+\frac{e^{\lambda }-1}{x}+\lambda '=0\,,\\
&&A_0''+A_0' \left(\frac{2}{x}-x (q A_0+1)^2 e^{\lambda -v}\right)-2 q e^{\lambda } (q A_0+1)=0\,,
\end{eqnarray}
where $x$ is the dimensionless radius defined in the same way as in Sec.~\ref{sec:neutral} and the prime now denotes the derivative with respect to $x$. The equations above are integrated for given values of $(v(0),A_0(0))$, such that the functions $v(r)$ and $A_0(r)$ smoothly matches the exterior Reissner-Nordstr\"om spacetime at the star radius. In this coordinate system, $\lambda(r)$ is not continuous across the sharp surface, similarly to the uncharged case.

We have performed the integration, investigating different values of the charge $q$. The result can be seen in Fig.~\ref{fig:mass-radius-charged}, where we plot the mass-radius curves for these charged stars. There are two families of solutions. The first branch does not have a black hole limit, and are the solutions with a maximum mass in Fig.~\ref{fig:mass-radius-charged}.
The maximum mass for each curve increases with $q$, as in the case with no self-interactions~\cite{Jetzer:1989av}. The configurations shown in the figure starting from the left part of the diagram have as limiting case the one presented in Sec.~\ref{sec:neutral} for $q=0$. Since these solutions are charged, their radius can be smaller than $2{\cal M}$, without becoming a BH. As such, we compare the stellar radii with the horizon or the photonsphere radii of a Reissner-Nordstr\"om BH with the same mass and charge, located at
\begin{equation}
r_+={\cal M}+\sqrt{{\cal M}^2-Q^2}, ~~	R_{\rm photon}=\frac{1}{2}(3{\cal M}+\sqrt{9{\cal M}^2-8Q^2})\,,\label{radius_charged}
\end{equation}
respectively. Our solutions always obey the Reissner-Nordstr\"om bound, i.e., $Q<{\cal M}$. Therefore, we can use $R/r_+$ and $R/R_{\rm photon}$ to analyze the compactness of charged BSs.

\begin{figure}
	\centering
	\includegraphics[width=0.48\linewidth]{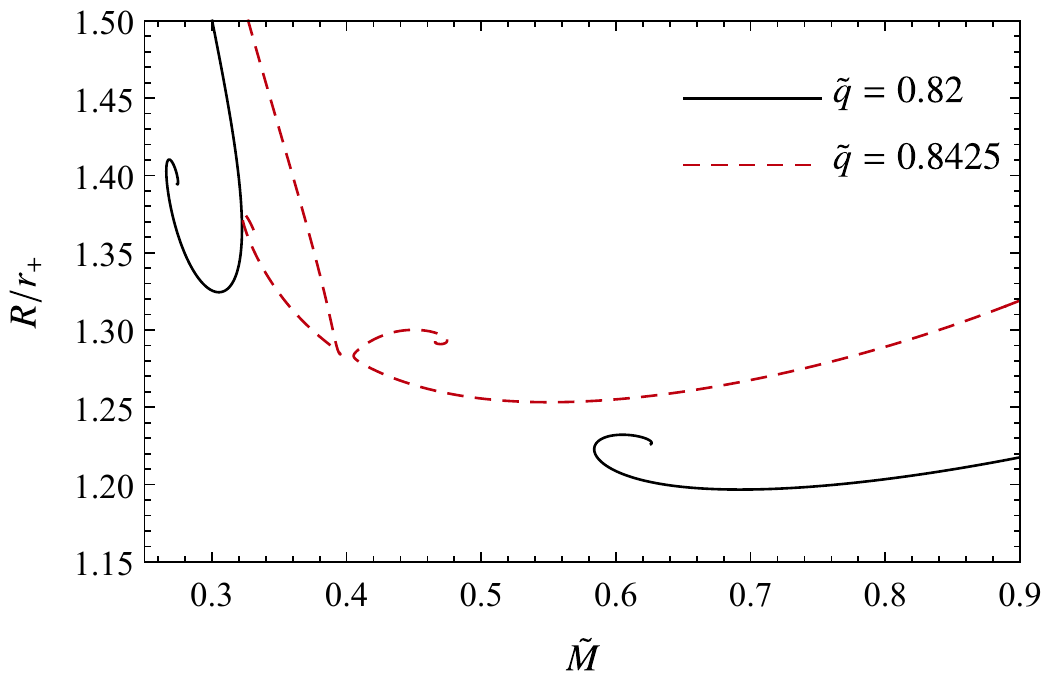}
	\includegraphics[width=0.48\linewidth]{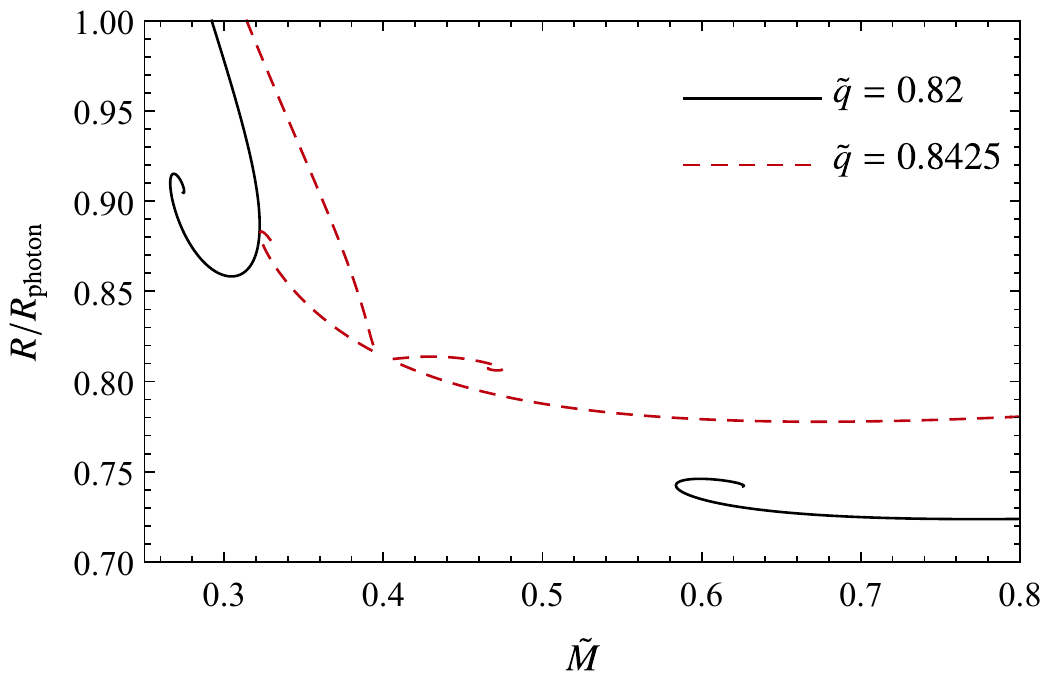}\\
	\includegraphics[width=0.48\linewidth]{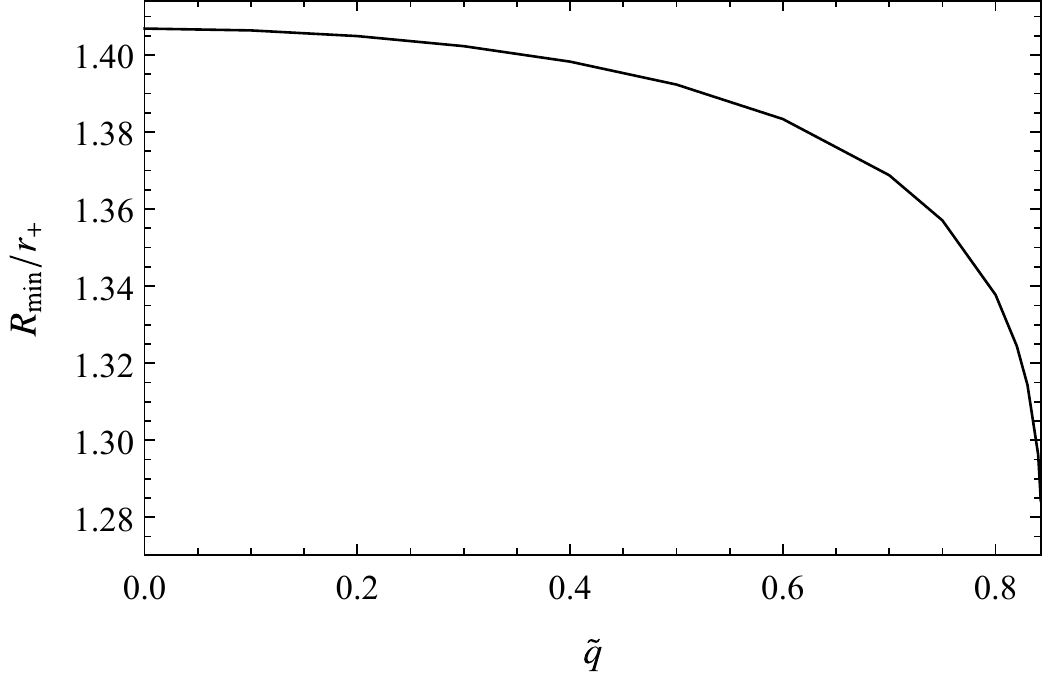}
	\includegraphics[width=0.48\linewidth]{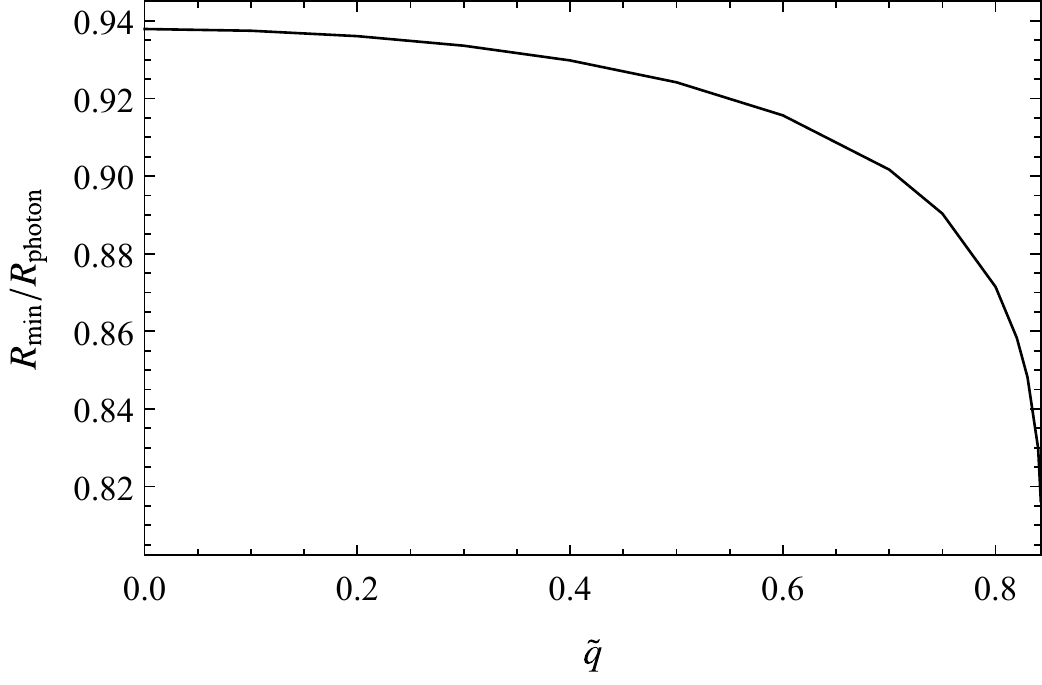}
	\caption{Top panels: Radius normalized by $r_+$ (top-left) and by the light-ring (top-right) radius, as defined in Eq.~(\ref{radius_charged}). We also show the radius for the second branch of solutions. Bottom panels: Minimum normalized radius as function of the charge of the scalar field, for the branch that recovers the uncharged solutions. We can see that the charge enables more compact solutions.}
	\label{fig:min_rad}
\end{figure}
We also see an additional branch of solutions -- in the top-right region in Fig.~\ref{fig:mass-radius-charged} -- which only exists in the charged case. We show solutions only for the case $\tilde q=0.8425$, but we also find this second branch for smaller values of $q$. We picked this value of charge because for higher we do not find BS solutions. In addition, while the charged configurations shown in Fig.~\ref{fig:mass-radius-charged} do respect the bound $Q<{\cal M}$, we also found solutions with $Q>{\cal M}$ in the second branch. 
In other words, the nonlinearities allow for higher charges. Such a property had been observed before as well~\cite{Herdeiro:2020xmb}.

In the top panels of Fig.~\ref{fig:min_rad} we show the radius of these configurations, normalized by $r_+$ (left) and the light-ring (right) radius $R_{\rm photon}$ for the two branches, considering $\tilde q=0.82$ and $0.8425$. The family of solutions in the left part of the plots are the same as the one shown in Fig.~\ref{fig:mass-radius-charged}, i.e., the ones that reduce to the uncharged case in the limit $q\to 0$. In the bottom panel of Fig.~\ref{fig:min_rad} we see that in the charged case the solution comes much closer to $r_+$. 

\subsection{Beyond the thin-wall approximation}
In general, one can also integrate the full system of equations, given by (\ref{eq:einl})--(\ref{eq:einphi}). This task, however, demands a highly precise and accurate numerical computation of the eigenvalue problem, especially for the higher compactness solutions with solitonic potential. For the uncharged case ($q=0$), the BS solutions of the solitonic potential has been obtained in a number of previous works~\cite{Macedo:2013jja,Cardoso:2016oxy}. As far as we are aware, however, the solitonic potential in the charged case has not been as explored, although many works analyzed self-interacting cases before (e.g., ~\cite{Kleihaus:2009kr,Pugliese:2013gsa,Kan:2017rqk,Brihaye:2009dx,Brihaye:2014gua,Kichakova:2013sza,Collodel:2019ohy}). Here we present some results in this direction.

Once, again, we use the following potential
\begin{equation}
V_S=\mu^2|\Psi|^2\left(1-\frac{|\Psi|^2}{\sigma^2}\right)^2,
\end{equation}
where the false vacuum, once again, is given by $\phi_0\sim\sigma$. For high values of $\sigma$, the case of massive scalar fields is recovered. For small values of $\sigma$, where the thin-wall approximation can be applied, the system can generate highly compact configuration, as can be seen below.

To integrate the solutions numerically, we proceed as follows. We eliminate the frequency $\omega$ by redefining the vector field, introducing $\tilde{A}(r)=qA_0(r)+\omega$. We require that the solution is regular at the origin, where the scalar field is $\phi_0(r=0)\equiv \phi_c$, and that the scalar field decays exponentially for $r\gg\mu^{-1}$. Therefore, the spacetime is asymptotically Reissner-Nordstr\"om. In summary, the boundary conditions at the origin depends on two parameters, namely $(v_c,\tilde{A}(0))$, one of which can be set to unity through a rescaling in the equations. The remaining parameter is found by a shooting procedure from the two-point boundary value problem (regularity at the origin and asymptotic flatness).

We explore BS solutions for different values of $(\sigma,q)$. There is a critical value of the charge $q$ of the field beyond which the solitonic solutions cannot exist, due to electromagnetic repulsion~\cite{Jetzer:1989av}. Our computations show that the critical charge depends on the nonlinearities of the potential. We also see indications for highly compact solutions for non-negligible charge, being more compact than the uncharged case. This confirms the predictions of the thin-wall approximation, which shows that the field charge increases the maximum compactness of BSs.

\begin{figure}
	\centering\includegraphics[width=0.49\linewidth]{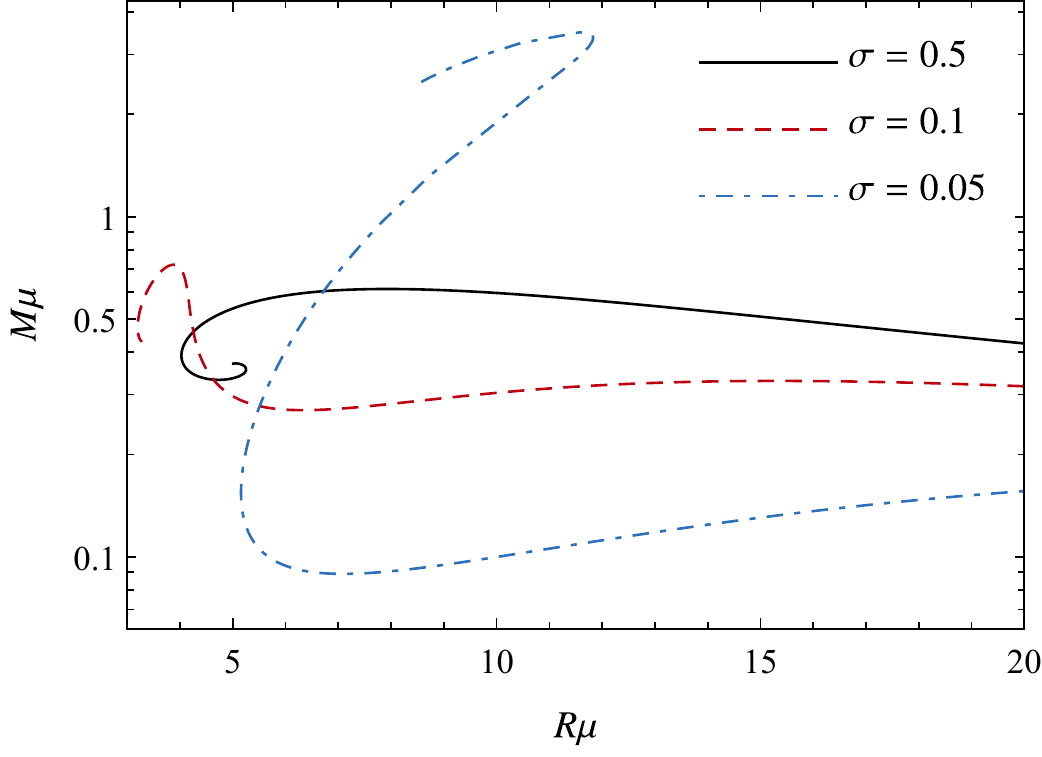}
	\includegraphics[width=0.49\linewidth]{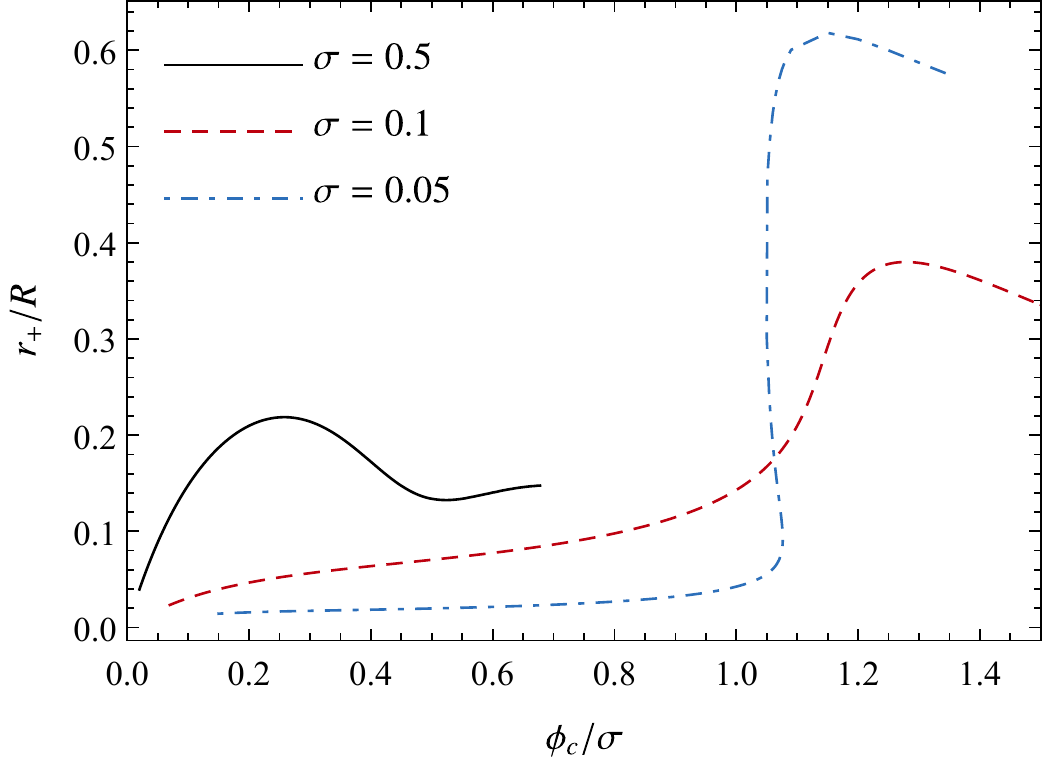}
	\caption{Left panel. Mass-radius relation for solitonic BS with $q/\mu=0.1$ and different values of $\sigma$. Right panel. Compactness for the same configurations of the left panel, as function of the central value of the scalar field.}
	\label{fig:mr_q1}
\end{figure}

\begin{figure}
	\centering\includegraphics[width=.49\linewidth]{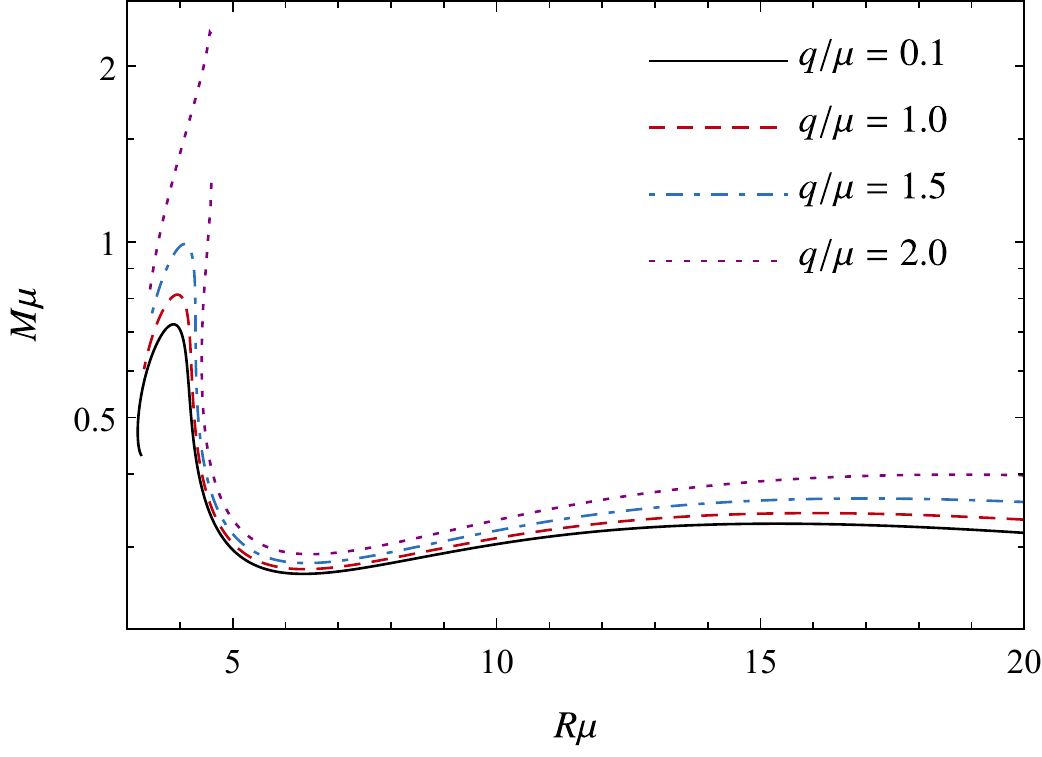}
	\includegraphics[width=.49\linewidth]{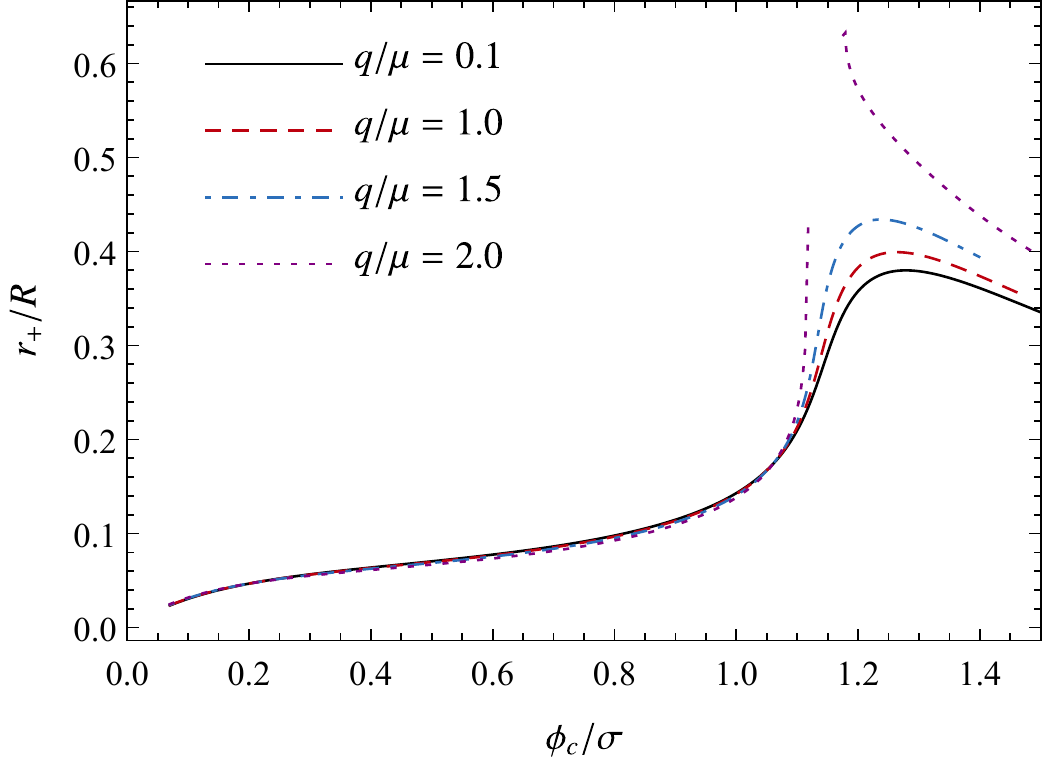}
	\caption{Mass-radius relations for solitonic charged BSs, considering different values for the charge of the scalar field $q$. For the case $q/\mu=2.0$, there is a discontinuity in the solution and the most compact have $Q\to M$. The most compact configuration has radius of $R\approx 0.944 R_{\rm photon}$.}
	\label{fig:higherq}
\end{figure}

\begin{figure}
	\centering\includegraphics[width=.5\linewidth]{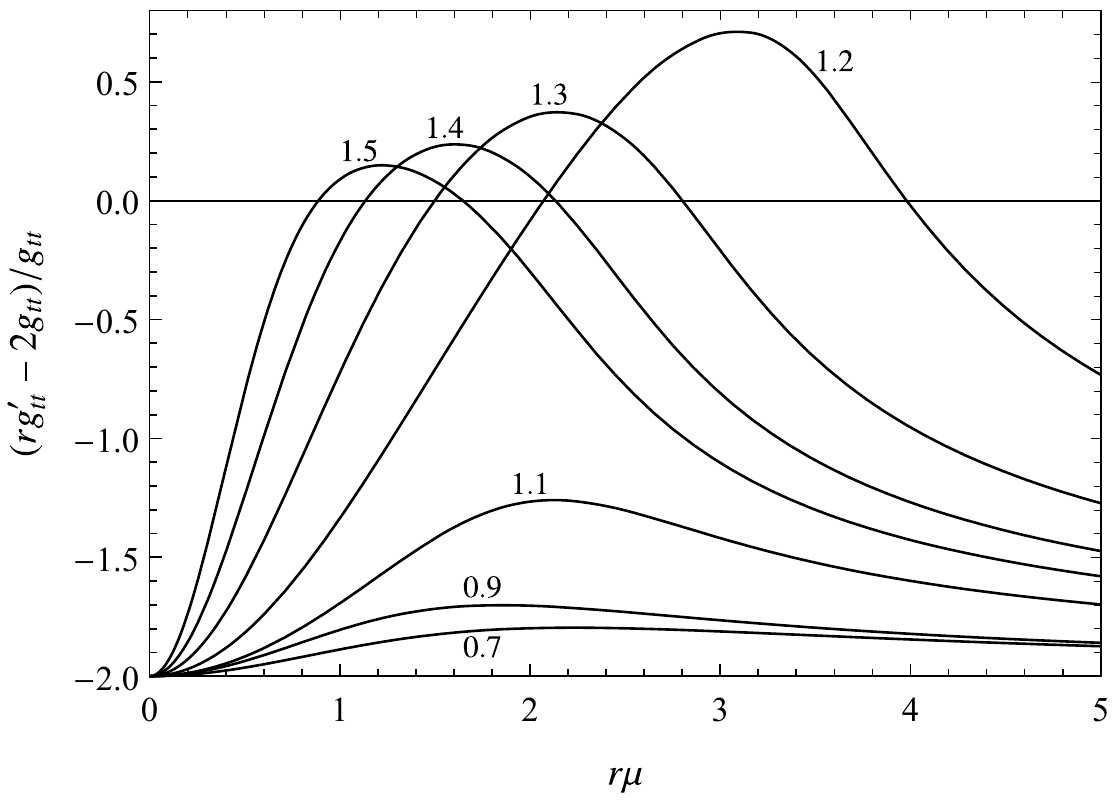}
	\caption{This figure quantifies the existence of light rings, by showing the discriminator (\ref{eq:light}) as function of $r$, for $q/\mu=2$ and $\sigma=0.1$ (purple dotted curve in Fig.~\ref{fig:higherq}). Light-rings exist at the position the curves cross the horizontal axis. Each curve corresponds to a different central scalar field $\phi_c/\sigma$, designated by the numbers next to the curves.}\label{fig:light}
\end{figure}

\begin{figure}
	\centering\includegraphics[width=1\linewidth]{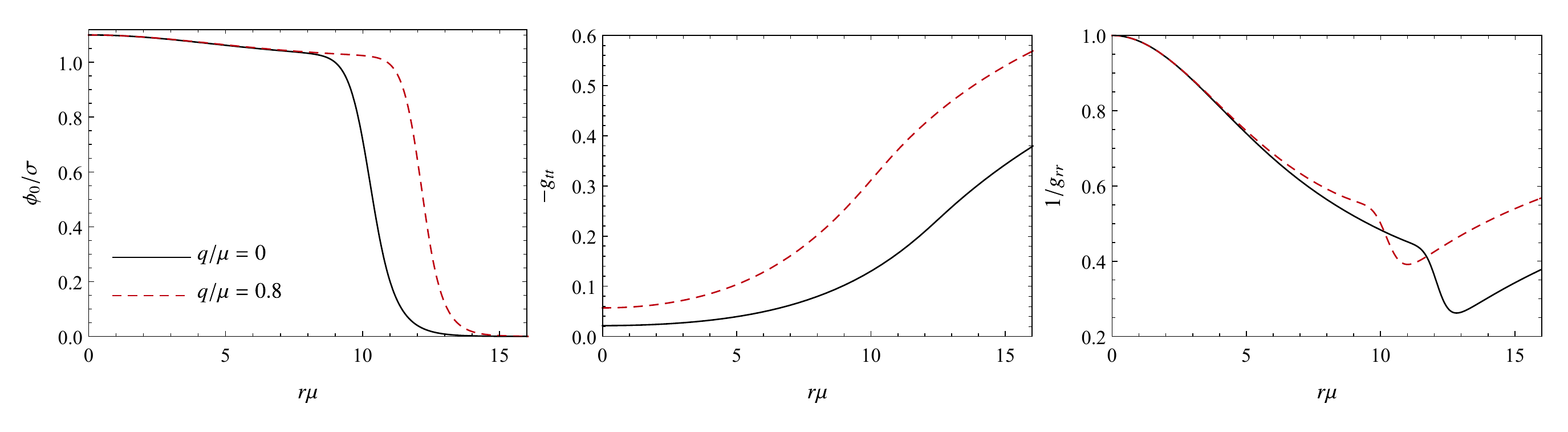}
	\caption{Scalar field and metric profiles for uncharged and charged ($q/mu=0.8$) BS configurations. For the uncharged case, we have $(R/M,R/R_{\rm photon})=(3.293,1.098)$ and for charged $(Q/M,R/M,R/R_{\rm photon})=(0.805,2.325,0.939)$.}
	\label{fig:background}
\end{figure}

In Fig.~\ref{fig:mr_q1} we plot the mass-radius relation considering small field charge $q/\mu=0.1$. As mentioned before, for high values of $\sigma$ we recover the standard results from considering only the mass term in the potential. As we decrease the value of $\sigma$, higher masses and compactness can be achieved at $\phi_c\sim\sigma$, with the mass radius curves behaving more similarly to the thin-wall approximation. Therefore, the small $\sigma$ limit corresponds to the regime in which the thin-wall is valid.

We now investigate the influence of the field charge in the solutions. We focus on the $\sigma=0.1$ case, for simplicity. For lower values of $\sigma$, the solutions are increasingly hard to find. We show our results in Fig.~\ref{fig:higherq}. Going to higher values of the field charge the solutions becomes more and more compact. We note a discontinuity for the $q/\mu=2.0$ case. By investigating the metric functions at the origin near the discontinuity, we see that  $g_{tt}$ approaches zero in that region. This indicates that we are possibly in a region of parameter space for which there is a small black hole at the center of a bosonic cloud. Hairy BHs exist in this theory, and were explored before in the literature~\cite{Hong:2020miv,Herdeiro:2020xmb}. These solutions indicate that we can indeed generate very compact solutions from charged BS configurations.

Because there are two possible BH solutions in the charged scalar field theory (Reissner-Nordstr\"om BHs and a BH carrying also scalar hair) a note of caution is needed for the compactness definition. As mentioned above, we are comparing these stars with the Reissner-Nordstr\"om BHs with the same parameters, analyzing their compactness according to this comparison. However, as indicated by the above paragraph, we can be closer to a hairy BH and perhaps be actually computing the size of a scalar cloud.

The solutions explored in Fig.~\ref{fig:higherq} have radii larger than the corresponding event horizon and light-ring of a Reissner-Nordstr\"om spacetime. But from the above reasoning, these stars might be connected to a different type of BH. It is, therefore, interesting to investigate the existence of light-ring in charged boson star cases. Light-rings can be obtained by looking for the zeroes of~\cite{Cardoso:2008bp,Cardoso:2014sna}
\begin{equation}
\frac{r g_ {tt}'(r)-2g_{tt}(r)}{g_{tt}(r)}\label{eq:light}\,.
\end{equation}
In Fig.~\ref{fig:light} we show such quantity for $q/\mu=2$ and $\sigma=0.1$. We show different configurations, corresponding to different central fields. We clearly see that light-rings can appear for some solutions (in pairs, one unstable the other stable).

Finally, to illustrate that the thin-wall approximation can be used in the charged BS configurations as well, in Fig.~\ref{fig:background} we show the scalar field and metric functions for a charged and uncharged BS configuration, for $\sigma=0.1$. We note that the scalar field behaves approximately as a step function in both cases. Also, we point out that, while the $g_{tt}$ is very smooth, we see that $g_{rr}$ has dent at the surface, which translate to the discontinuity in the thin-wall approximation.
\section{Discussion and conclusion} \label{sec:conclusion}
There is an important effort to understand the limits of compact solutions and how they can challenge the BH paradigm. In this paper, we studied self-gravitating fundamental bosonic fields. In particular, we focused mostly in configurations for which the thin wall approximation can be performed, but we also computed full numerical solutions to strengthen our understanding of the spectrum of solutions. Our results show that spherically symmetric stars formed by bosonic scalar field configurations can potentially be very compact, with a radius slightly smaller than the corresponding light-ring. In particular, the light-ring seems to play an important role when higher dimensions are considered, with the minimum radius of the star converging to the light-ring as $d\to\infty$. Our results also indicate that it is very difficult to get bosonic stars with a surface parametrically close to the Schwarazschild radius, or analogs thereof for charged configurations.

We also computed vector BSs configurations, including up to sixth-order terms in the self-interacting potential. This case allows for solitons even in flat spacetime solutions. We note, however, that differently from the scalar case, we found a maximum compactness of around ${\cal C}_{\rm max}\sim 0.36$ and, as such, these stars do not present light-rings. It would be interesting to analyze self-interacting potentials for the vector BSs that remove possible divergences in the solutions, generating, therefore, solutions for generic values of the central field/density. Such study is beyond the scope of this work, however. It is also interesting to consider different coupled fields and to understand how they impact on the maximum compactness. 

We also studied charged scalar fields. In the thin-wall approximation, the compactness is increased when compared to the uncharged case. The full numerical construction confirms these findings, that indeed the scalar field's charge increase the compactness of the solutions, and also shows the existence of light-rings in some of these solutions. It would be interesting and important to perform a more thorough analysis of these stars, especially in light of new hairy black hole solutions in this theory \cite{Herdeiro:2020xmb,Hong:2020miv}. In particular, it is important to understand how these stars are connected to the black hole branch, but we leave this for future work.

\section*{Acknowledgements}
%
We are indebted to Waseda University for warm hospitality while this work was being completed.
V. C. is a Villum Investigator supported by VILLUM FONDEN (grant no. 37766) and a DNRF Chair supported by the Danish National Research Foundation.
V.~C.\ acknowledges financial support provided under the European Union's H2020 ERC 
Consolidator Grant ``Matter and strong-field gravity: New frontiers in Einstein's 
theory'' grant agreement no. MaGRaTh--646597.
C.F.B.M. would like to thank Fundação Amazônia de Amparo a Estudos e Pesquisas (FAPESPA), Conselho Nacional de Desenvolvimento Científico e Tecnológico (CNPq), and Coordenação de Aperfeiçoamento de Pessoal de Nível Superior (CAPES), from Brazil.
This work was supported in part by JSPS KAKENHI Grant Numbers JP20J12436, JP17H06359,  JP19K03857 and JP19H01895 and by Waseda University Grant for Special Research Projects(Project Number: 2019C-640).
This project has received funding from the European Union's Horizon 2020 research and innovation 
programme under the Marie Sklodowska-Curie grant agreement No 101007855.
We thank FCT for financial support through Project~No.~UIDB/00099/2020.
We acknowledge financial support provided by FCT/Portugal through grants PTDC/MAT-APL/30043/2017 and PTDC/FIS-AST/7002/2020.
The authors would like to acknowledge networking support by the GWverse COST Action 
CA16104, ``Black holes, gravitational waves and fundamental physics.''

\newpage

\appendix

\section{Solution of Eq. (\ref{ode_solitonic})}
We solve Eq.  (\ref{ode_solitonic}) by the power-series expansion.
From the boundary condition at the origin, we expect that the solution is even function of $x$.
Hence we expand the metric function $e^v$ as
\be
e^v=\sum_{n=0}^\infty a_n x^{2n}\,.
\label{power_law}
\ee

This gives two branches of the expansion coefficients $a_n$: 
\\
(i)
\beq
a_0&= &0
\\
a_1&=&{2\over d-3}
\\
a_n&=&0~~(n\geq 2)
\eeq
(ii)
\beq
a_0&= &e^{v_c}
\\
a_1&=&{2\over d-1}
\\
a_2&=&{4\over (d^2-1)(d-2)}e^{-v_c}
\\
a_3&=&-{16(d-4)(d^2-5d+2)\over (d-3)(d-2)^2(d-1)^2(d+1)(d+3) }e^{-2v_c}
\\
a_4&=&{32(d-5)(d-9d+6)(d^3-6d^2+9d+4)\over (d-3)^2(d-2)^3(d-1)^3(d+1)(d+3)(d+5) }e^{-3v_c}
\\
\cdots
\,,
\eeq
where $v_c$ is the value of $v$ at the origin.

The first one (i) gives the exact solution
\be
e^v={2x^2\over d-3}\,.
\ee
It is singular at the origin, but gives the critical compactness at any radius as
\be
{\cal C}=\frac{xv'}{d-3+xv'}={2\over d-1}\,.
\ee
The metric function $e^\lambda$ is obtained by Eq. (\ref{eq_lambda}) as
\be
e^\lambda={(d-2)(d-3)+2x^2 e^{-v}\over (d-2)(d-3)+(d-2)xv'}
\ee
The solution (i) gives
\be
e^\lambda={d-3\over d-2}\,.
\ee

The second solution (ii) is regular at the origin, but becomes  infinite series.
However the expansion coefficient $a_n$ is proportional to $1/d^{n+1}$ in the limit of $d\rightarrow \infty$. As a result, we find the asymptotic solution
\be
e^v\approx e^{v_c}+{2x^2\over d-1}+\cdots
\,,
\ee
which gives the compactness as
\be
{\cal C}\approx {2\over (d-1)\left[1+{(d-3)e^{v_c}\over 2x_S^2}\right]}\leq {2\over d-1}
\,,
\ee
where $x_S$ is the surface radius.
The metric function $e^\lambda$ for the solution (ii)  is given by
\be
e^\lambda\approx {(d-1)(d-2)(d-3)e^{v_c}+2(d^2-4d+5)x^2\over
	(d-1)(d-2)(d-3)e^{v_c}+2(d-1)(d-2)x^2}\leq 1\,.
\ee

\section*{References}
\bibliographystyle{iopart-num}
\bibliography{References}

\end{document}